\documentclass[twocolumn,showpacs,preprintnumbers,aps,amsmath,amssymb]{revtex4}

\usepackage{graphicx}
\usepackage{dcolumn}
\usepackage{bm}

\begin{document}

\title{Superconducting Gap Structure of LaFePO Studied by Thermal Conductivity}

\author{M.~Yamashita,$^{1,\ast}$ N.~Nakata,$^1$ Y.~Senshu,$^{1}$ S.~Tonegawa,$^{1}$ K.~Ikada,$^1$ K.~Hashimoto,$^{1}$ H.~Sugawara,$^2$ T.~Shibauchi,$^1$ and Y.~Matsuda$^1$}

\affiliation{
$^1$Department of Physics, Graduate School of Science, Kyoto University, Kyoto 606-8502, Japan \\
$^2$Faculty of Integrated Arts and Sciences, The University of Tokushima, Tokushima 770-8502
}

\date{\today}


\begin{abstract}
The superconducting gap structure of LaFePO ($T_c=7.4~$K) is studied by thermal conductivity ($\kappa$) at low temperatures in fields $H$ parallel and perpendicular to the $c$ axis. 
A clear two-step field dependence of $\kappa(H)$ with a characteristic field $H_s(\sim 350$~Oe) much lower than the upper critical field $H_{c2}$ is observed.
In spite of large anisotropy of $H_{c2}$, $\kappa(H)$ in both $H$-directions is nearly identical below $H_s$.
Above $H_s$, $\kappa(H)$ grows gradually with $H$ with a convex curvature, followed by a steep increase with strong upward curvature near $H_{c2}$. 
These results indicate the multigap superconductivity with active two-dimensional (2D) and passive 3D bands having contrasting gap values. 
Together with the recent penetration depth results, we suggest that the 2D bands consist of nodal and nodeless ones, consistent with the extended $s$-wave symmetry.

\end{abstract}


\maketitle

Recent discovery of a new class of Fe-based superconductors \cite{kam08} has attracted much attention. 
Among them,  FeAs-based compounds have aroused great interest because of the high transition temperature $T_c$.  Undoped arsenide LaFeAsO(La-1111) is non-superconducting and has a spin-density-wave (SDW) ground state, but becomes superconducting ($T_c=25$~K) when electrons are doped \cite{lue09}. By changing the rare-earth ion, $T_c$ reaches as high as 55~K in SmFeAs(O,F) \cite{ren08}. A key question is the origin of the pairing interaction.  Since the symmetry of the superconducting order parameter is intimately related to the pairing interaction at the microscopic level, its identification is of primary importance.

Fully gapped superconducting states in FeAs-based superconductors have been reported by the penetration depth measurements of Pr-1111 \cite{has09}, Sm-1111 \cite{mal08}, and Ba$_{1-x}$K$_x$Fe$_2$As$_2$ (Ba-122) \cite{has09b}, angle-resolved photoemission 
\cite{din08}, thermal conductivity \cite{luo09} and NMR relaxation rate \cite{yas09} measurements of Ba-122.  Some of them give evidence of multiband superconductivity with two distinct gaps. On the other hand, the NMR of La-1111 \cite{nak08} and Pr-1111 \cite{mat08} and the penetration depth measurements of Ba(Fe$_{1-x}$Co$_x$)$_2$As$_2$ \cite{gor09} suggest the presence of low-lying excitations, which could be indicative of nodes. Theoretically, it is proposed that a good nesting between hole and electron pockets prefers the ``$s_\pm$" symmetry where the gap is finite at all Fermi surfaces but changes its sign on different bands \cite{maz08}. 
Recent neutron resonant scattering \cite{chr08} and the impurity effects on the penetration depth \cite{has09b,shi09} are consistent with this symmetry.

The phosphide LaFePO isomorphic to LaFeAsO exhibits superconductivity at $T_c\sim 7$~K without doping \cite{kam06}.   In spite of the similarity in the overall electronic structure \cite{vil08}, the magnetic and superconducting properties of LaFePO and LaFeAsO are very different.  In fact, the undoped stoichiometric LaFePO is non-magnetic.   Recently, a superconducting state with well-developed line nodes in the gap function has been suggested in LaFePO by a linear temperature dependence of the penetration depth at low temperatures \cite{fle09}.  However, on which Fermi surfaces the nodes locate in the multiband electronic structure is not yet clarified, and  while several candidates have been theoretically proposed \cite{gra09,kuroki09}, the superconducting symmetry in LaFePO remains elusive.  Thus the clarification of the detailed gap structure of LaFePO is expected to provide important clues to the origins of magnetism and superconductivity of Fe-based compounds.

Here, to shed further light on the gap symmetry of LaFePO, we present the thermal conductivity measurements at low temperatures.  The thermal conductivity probes delocalized low-energy quasiparticle excitations and is an extremely sensitive probe of the anisotropy of the gap amplitude.  We provide strong evidence of the multigap superconductivity, in a more dramatic fashion than FeAs-based superconductors, with two very different gap values. 
We show that there are passive 3D bands and two kinds of active 2D bands; one is fully gapped and the nodes inferred from the penetration depth measurements \cite{fle09} are most likely on the other 2D bands. 
This is compatible with the extended $s$-wave (nodal $s_{\pm}$) symmetry for the gap structure of LaFePO.

Single crystals with typical dimensions of $0.8 \times 0.4 \times 0.05$~mm$^3$ were grown by a Sn-flux method \cite{sug08}.  We carefully removed the Sn-flux at the surface of the crystals by the diluted hydrochloric acid.  
The resistivity and susceptibility measurements show the sharp superconducting transition at $T_c=7.4$~K (determined by the midpoint of the resistive transition), which is slightly higher than the values reported by other groups \cite{fle09,ham08}. The thermal conductivity $\kappa$ was measured by a standard four-wire steady method for a heat current {\boldmath $q$} within the $ab$ plane.  

The temperature dependence of the in-plane resistivity $\rho$ in the zero field (inset of Fig.~1) depends on $T$ as $\rho=\rho_0+AT^2$ with $\rho_0= 4.9~\mu\Omega$\,cm and $A = 3.3 \times 10^{-3}~\mu\Omega$\,cm/K$^2$ below 50~K down to $T_c$.  The residual resistivity ratio ($RRR$) is 28.  Clear de Haas-van Alphen (dHvA) oscillations were observed in samples from the same batch with nearly the same $RRR$ value~\cite{sug08}.  The upper critical fields at $T\rightarrow 0$~K estimated by the resistivity measurements are $\mu_0H_{c2}^c=1.0$~T for {\boldmath $H$}$\parallel c$ axis and $\mu_0H_{c2}^{ab}=8.6$~T for {\boldmath $H$}$\parallel ab$ plane.  This large anisotropy of the upper critical field $H_{c2}$ indicates that the bands active for the superconductivity have a very anisotropic 2D electronic structure.

\begin{figure}[t]
\includegraphics[width = 0.95 \linewidth,keepaspectratio]{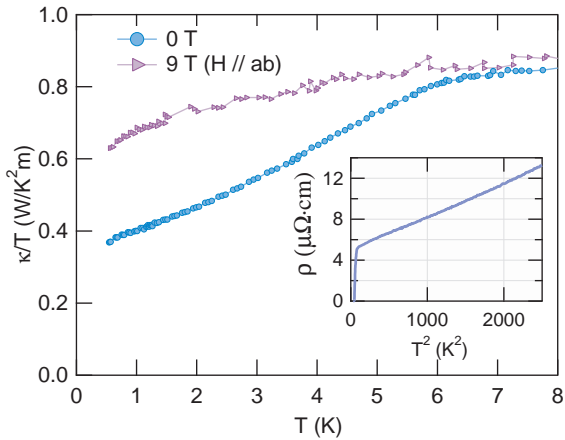}
\caption{\label{k_T}(color online)
Temperature dependence of $\kappa/T$ in zero field and at $\mu_0H=9$~T in the normal state above $H_{c2}^{ab}$ ({\boldmath $H$}$\parallel ab$, {\boldmath $H$}$\perp$ {\boldmath $q$}). 
Inset: $\rho$ plotted as a function of $T^2$.}
\end{figure}

Figure~\ref{k_T} depicts the temperature dependence of $\kappa/T$ in zero field and in the normal state above $H_{c2}^{ab}$.
As the temperature is lowered, $\kappa/T$ decreases below $T_c$.  
The electronic contribution in $\kappa$ at 1~K estimated by the Wiedemann-Franz law, $\kappa=L_0T/\rho$ ($L_0=2.44\times10^{-8}~\Omega$\,W/K is the Sommerfeld value),  is 0.5~W/K$^2$m, which is close to the observed value $0.65$~W/K$^2$m.  Thus the electron contribution dominates well the phonon heat contribution at least  below 1~K.

First we discuss the thermal conductivity in zero field.  A residual term $\kappa_{00}/T$ at $T \rightarrow 0$~K in $\kappa/T$ is clearly resolved.  In the nodal superconductors, such a residual term appears as a result of the impurity scattering which induces quasiparticles even at $T=0$~K.  In the presence of line nodes in a single band superconductor, $\kappa_{00}/T$ is roughly estimated as $\sim2(\xi_{ab}/\ell)\cdot (\kappa_n/T)$, where $\xi_{ab}$ is the in-plane coherence length, $\ell$ is the mean free path and $\kappa_n$ is the thermal conductivity in the normal state.  
Using $\ell=94$~nm from the dHvA measurements \cite{sug08} and $\xi_{ab}=\sqrt{\Phi_0/(2\pi H_{c2}^c)}=18$~nm, $\kappa_{00}/T$ is estimated to be $\sim$0.19~W/K$^2$m.  This value is comparable to the observed $\kappa_{00}/T\sim 0.30~$W/K$^2$m, but we note that this estimate includes large ambiguities due to the multiband effect which could alter effective $\xi_{ab}$ and $\kappa_n$. So this comparison alone cannot be taken as the evidence for line nodes in the superconducting gap. It should be also noted that the residual term may arise from an extrinsic origin, such as non-superconducting metallic region \cite{ham08} with high thermal conductivity, though the sharp superconducting transition and the observation of the dHvA oscillation indicate a good quality of the crystal.

More vital information on the gap structure can be provided by  field dependence of thermal conductivity at low temperatures. Field dependent part of $\kappa(H)$ in a mixed state mainly stems from the superconducting part of the crystals, even if a non-superconducting region was present in the crystal. 
Moreover, the phonon scattering at the low temperatures is governed by static defects and is therefore field independent. 

Figures~2(a) and (b) depict the field dependence of $\kappa(H)-\kappa(0)$ normalized by $\kappa(H_{c2})-\kappa(0)$ for {\boldmath $H$}$\parallel c$ and {\boldmath $H$}$\parallel ab$, respectively, measured at $T=0.46$~K (0.062$T_c$) by sweeping $H$ after zero-field cooling.  We note that little difference was observed  between the data measured in zero-field and field cooling conditions, indicating that the field trapping effect is very small.  For both field directions, the overall $H$-dependence of thermal conductivity is quite similar.  At very low fields, the thermal conductivity exhibits a pronounced increase with increasing $H$.  Remarkably, in spite of the large anisotropy of $H_{c2}$, $\kappa(H)$ is nearly identical for both $H$ directions at low fields and almost saturates at around $H_s\sim 350$~Oe, as shown in the inset of Fig.~2(b).  Above $H_s$, $\kappa(H)$ becomes anisotropic with respect to the field direction and is governed by the anisotropy of $H_{c2}$.  For both {\boldmath $H$}$\parallel c$ and {\boldmath $H$}$\parallel ab$, $\kappa(H)$ grows gradually with $H$ with a convex curvature, followed by a rapid increase with a concave curvature up to $H_{c2}$; there is an inflection point at $\sim H_{c2}/2$. 

\begin{figure}[t]
\includegraphics[width = 0.95 \linewidth,keepaspectratio]{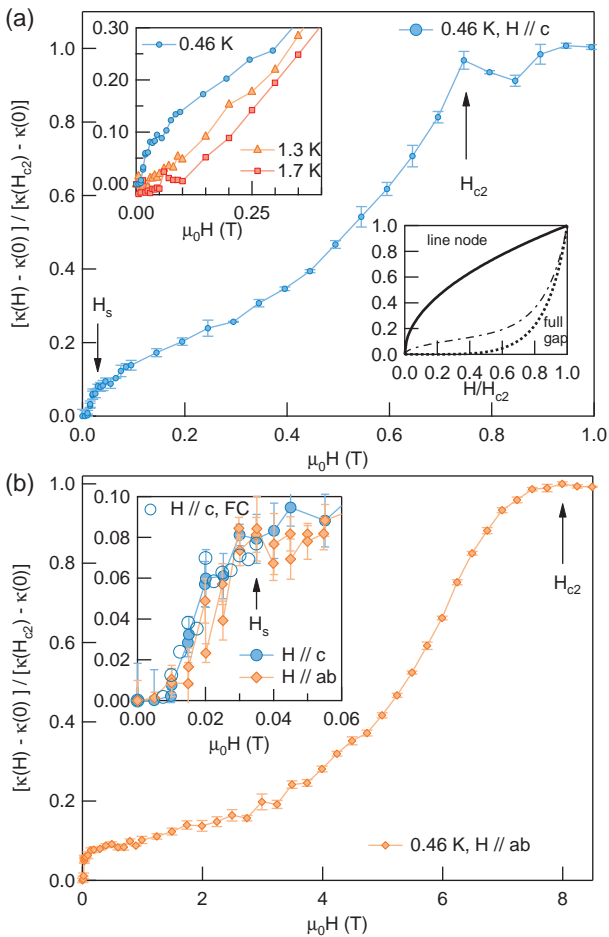}
\caption{\label{k_H_T}(color online)
(a) Field dependence of $\kappa(H)-\kappa(0)$ normalized by $\kappa(H_{c2}^{c})-\kappa(0)$ for {\boldmath $H$}$\parallel c$ at $T=0.46$~K. 
Upper inset: The same plot in low fields 
for {\boldmath $H$}$\parallel c$ at 0.46 K (circles), 1.3 K (triangles) and 1.7 K (squares).
Lower inset: Schematic field dependence of $\kappa(H)-\kappa(0)$ in superconductors with a full gap (dotted line), with line nodes (solid line), and with two kinds of gaps with and without nodes (dash-dotted line).
(b) Field dependence of $[\kappa(H)-\kappa(0)]/[\kappa(H_{c2}^{ab})-\kappa(0)]$ for {\boldmath $H$}$\parallel ab$ plane at 0.46~K.
Inset: A comparison of the low-field data for {\boldmath $H$}$\parallel ab$ with those for {\boldmath $H$}$\parallel c$ measured in the zero-field (filled circles) and field cooling (FC) conditions (open circles).}
\end{figure}

In superconducting states, the thermal transport is governed by delocalized quasiparticles, which shows a contrasting field dependence between fully-gapped and nodal superconductors, as illustrated in the lower inset of Fig.~2(a). In fully-gapped superconductors, the only quasiparticles states present at $T\ll T_c$ are those bound to the vortex cores and unable to transport heat, until these quasiparticle states within vortices begin to overlap with those within the neighboring vortices. Consequently, $\kappa(H)$ shows an exponential behavior with very slow grows with $H$ in low fields and a rapid increase near $H_{c2}$ (the dotted line). Such a field dependence is indeed reported in fully-gapped superconductors such as Nb \cite{low70} and BaNi$_2$As$_2$ \cite{kur09}.  In sharp contrast, the heat transport in superconductors with nodes or with a large anisotropy in the gap is dominated by contributions from delocalized quasiparticles outside vortex cores.  The most remarkable effect on thermal transport is the Doppler shift of the energy of quasiparticles \cite{vol93,mat06}.  In the presence of line nodes where the density of states has a linear energy dependence, $\kappa(H)$ increases in proportion to $\sqrt{H}$.  Then $\kappa(H)$ increases rapidly as soon as the field exceeds the lower critical field $H_{c1}$ \cite{mat06,pro02} (the solid line). If nodeless and nodal gaps are mixed in a multiband system without interband scatterings, an inflection point emerges in the intermediate field regime (the dash-dotted line).


\begin{figure}[t]
\includegraphics[width = 0.9 \linewidth,keepaspectratio]{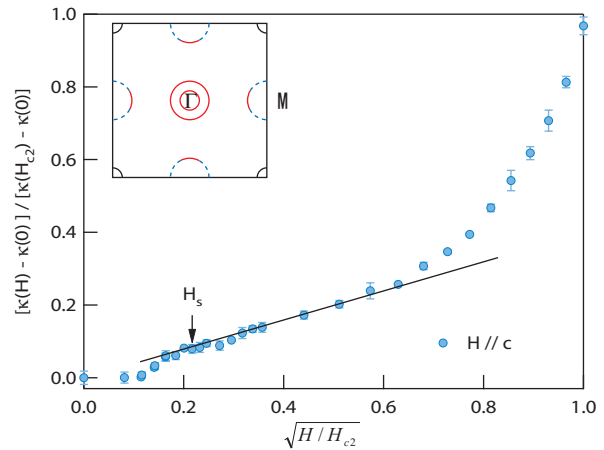}
\caption{\label{k_H}(color online)
The same data in the main panel of Fig.~2(a) plotted against $\sqrt{H/H_{c2}}$. The solid line is a guide for the eyes.
The inset illustrates the extended $s$-wave (nodal $s_\pm$) gap structure \cite{kuroki09} in the unfolded Brillouin zone. The sign of the gap changes between the solid and dotted lines.
The small band at the zone corner represents the 3D band around the $Z$ point.
}
\end{figure}

Now let us discuss the thermal conductivity of LaFePO below $H_s$ shown in the inset of Fig.~2(b). The nearly isotropic $\kappa(H)$ with respect to the field direction  definitely indicates that the steep increase of $\kappa(H)$ below $H_s$ is not due to the influence of the first vortex penetration field, which is expected to be much smaller than $H_{c1}\sim 100$~Oe, if the demagnetization is taken into account \cite{oka09}.  We can also rule out a possibility that the steep increase is caused by the remanent Sn-flux inside the crystal, because $\kappa(T)$, magnetization, and microwave surface impedance measurements \cite{ton09} show no anomaly at $T_c$ of Sn ($=3.72$~K).  Moreover, as shown in the upper inset of Fig.~2(a), the low-field steep increase disappears at $T\sim 1.5$~K well below $T_c$ of Sn. Thus, we conclude that the observed $H$-dependence of $\kappa(H)$ below $H_s$ is related to the quasiparticle excitations inherent to LaFePO.

The steep increase and subsequent gradual increase of $\kappa(H)$ above $H_s$ for both {\boldmath $H$}$\parallel c$ and {\boldmath $H$}$\parallel ab$ indicate that a substantial portion of the quasiparticles is already restored at $H_s$, much below $H_{c2}$.  Such a two-step field dependence has been reported in MgB$_2$ \cite{sol02}, NbSe$_2$ \cite{boa03}, PrOs$_4$Sb$_{12}$ \cite{sey05}, and URu$_2$Si$_2$ \cite{kas07}, providing direct evidence for the multiband superconductivity.  Here $H_s$ is interpreted as a ``virtual upper critical field" that controls the field dependence of the smaller gap of the ``passive" band.  Its superconductivity is  most likely induced by the proximity effect of the ``active" bands with primary gap.  The ratio of the large and small gaps is roughly estimated to be $\Delta_L/\Delta_S\sim \sqrt{H_{c2}^c/H_s}\sim 6$.  If we take $\Delta_L\sim 1.7~k_BT_c\sim13$~K, we obtain $\Delta_S\sim 2$~K. This is in good correspondence with the fact that the steep increase in $\kappa(H)$ below $H_s$ is smeared by a similar temperature scale ($\gtrsim 1.5$~K), as shown in the upper inset of Fig.~2(a).

The observed multiband superconductivity is compatible with the band structure of LaFePO, which has been examined theoretically and experimentally.  The band structure calculations show that Fermi surface consists of two electronic cylinders centered at the $M$ point and two hole cylinders centered at the $\Gamma$ point, together with a single hole pocket with 3D like dispersion at the $Z$ point in the Brillouin zone (see the sketch in Fig.~3) \cite{vil08,col08}. 
The 3D band is suggested to appear in LaFePO, not in LaFeAsO, and has a character of the $3d_{3z^2-r^2}$ orbital which is expected to have a weak coupling to other 2D bands \cite{vil08}. 
Note that only the 2D cylindrical electron and hole bands have been reported by photoemission \cite{lu08} and dHvA measurements \cite{sug08,col08}.

Nearly isotropic $\kappa(H)$ with respect to the field direction below $H_s$ shown in the inset of Fig.~2(b) indicates that the smaller gap is present most likely in the 3D hole pocket. This is consistent with the expected weak coupling between the 3D and 2D bands.  This passive 3D band is inferred to be fully gapped because of the following reasons.   The field dependence of $\kappa$ below $H_s$ does not show strong $\sqrt{H}$ dependence expected for line nodes.  Moreover, since the coherence length of the smaller gap, $\xi_s = \sqrt{\phi_0/(2\pi H_s)}\simeq100$~nm, is comparable to the mean free path, it is unlikely that a nodal superconductivity can survive against such a ``dirty" condition ($\xi_s \simeq \ell$).

Next we discuss the gap structure of the 2D bands from $\kappa(H)$ above $H_s$, where essentially all quasiparticles of the 3D band with smaller gap have already contributed to the heat transport.  As shown in Fig.~3,  $\kappa(H)$ for {\boldmath $H$}$\parallel c$  increases as $\sim\sqrt{H}$ just above $H_s$ to $\sim 0.4 H_{c2}$ (the $\sqrt{H}$-dependence is not clear below $H_s$ in our resolution).  This $H$-dependence and the appearance of the inflection point from convex to concave $H$-dependence are in sharp contrast to the $H$-dependence of the simple fully-gapped superconductors, in which $H$-dependence is always concave well below $T_c$~\cite{low70}.    Such a convex (sub-linear) $H$-dependence at low fields appears when the gap is highly anisotropic with a large amplitude modulation.  On the other hand, the concave $H$-dependence just below $H_{c2}$ has never been reported in  superconductors with large anisotropic gap, such as Tl$_2$Ba$_2$CuO$_{6+\delta}$ \cite{pro02}, CePt$_3$Si \cite{iza05}, UPt$_3$ \cite{sud97} and LuNi$_2$B$_2$C \cite{boa01}.  Therefore, it is likely that at least one of the active 2D bands is fully gapped without nodes, although more detailed theoretical investigations of the multiband systems are required \cite{Vekpr}.  In fact, the $H$-dependence of $\kappa(H)$ with two kinds of gaps with and without nodes in the absence of interband scatterings shows an inflection behavior (see the dash-dotted line in the lower inset of Fig.~2(a)), which qualitatively reproduces the data. This result, along with the finite $\kappa_{00}/T$ observed in our dHvA-available clean crystal, supports the nodal superconductivity suggested by the linear temperature dependence of the superfluid density \cite{fle09}.  Thus the whole $H$-dependence of $\kappa$ above $H_s$ implies that the 2D bands consist of two kinds; one has nodes and the other is fully gapped. 


Finally we discuss the position of the nodes. 
As candidates of the gap structure with line nodes, the ``nodal $s_\pm$-wave" and ``$d$-wave" symmetries have been proposed  for LaFePO \cite{kuroki09}.
We infer that the $d$-wave can be excluded because it possesses line nodes in the 3D band (as well as the 2D hole bands), which is unlikely for the reasons discussed above. The nodal $s_\pm$-wave structure has a nodal gap on the 2D electron band around $M$ point and the 2D and 3D hole bands are fully gapped (see the sketch in Fig.~3). The gap size of the 2D electron and hole bands can be comparable to each other \cite{kuroki09}. 
Therefore, we suggest that the nodal $s_\pm$-wave structure can be the best candidate for the gap symmetry of LaFePO.

In summary, from the measurements of the thermal conductivity,  LaFePO is found to be a multigap superconductor with 2D active and 3D passive bands. The peculiar field dependence of $\kappa$ provides a stringent constraint on the superconducting gap structure in this system: there exist fully gapped 2D and 3D bands and the nodes locate most likely on the other 2D bands. These results are consistent with the nodal $s_\pm$-wave symmetry proposed for the superconducting state of LaFePO.

We thank R. Arita, A. Carrrington, H. Ikeda, K. Kontani, K. Kuroki, and I. Vekhter for valuable discussion. This work was supported by KAKENHI from JSPS.


\end{document}